# Transformation of polarons to bipolarons in disordered matter

I. Sakellis, A. N. Papathanassiou [a)], and J. Grammatikakis

University of Athens, Department of Physics, Section of Solid State Physics, Panepistimiopolis, GR 15784 Zografos, Athens, Greece

The polaron to bipolaron transition in disordered media under the influence of a broadband ac electric field is suggested: ac conductivity vs. frequency measurements in disordered media with inhomogeneous disorder induce spatio-temporal fluctuations of the density of polarons yielding polaron to bipolaron transformation. The external field results in the transition, alternatively to an increasing doping level. The assertion is confirmed by performing successive complex impedance measurements in disordered media. A systematic increase of the real part of the ac conductivity in the frequency domain, followed by mutual reduction of the magnetic inductivity of conducting polypyrrole, are explained.

[a)] Corresponding author; e-mail address: antpapa@phys.uoa.gr

Since the discovery of the metal to insulator transition by doping in conjugated polymers [1, 2], considerable experimental investigation on the electric charge transport in conducting polymers has been reported [3, 4]. Extensive efforts were made to achieve a deep theoretical understanding of transport phenomena. The objectives is the construction of a generalized frame for electrical conductivity in conducting polymers [5, 6] and the preparation of well characterized metallic or semi-conducting polymers for many different technological applications [7, 8]. Polymer chains of various lengths, with conformational disorder and random orientation constitute a polymer network.

Conducting polypyrrole is characterized by inhomogeneous disorder: short – range ordering in the polymer chains result in the appearance of conducting grains embedded into an insulating amorphous matrix [3, 4, 9-13]. An electric charge carrier (such as a polaron) can hop along each chain (intra-chain transfer) and over cross-linked chain clusters. Charge flow is favored within the ordered polaron-rich grains, while inter-grain transport occurs by penetrating the highly disordered insulating matrix [14]. The system can be virtually described as a network of conducting pathways with distributed effective length [15].

It is an experimental fact [16 – 22] that the ac conductivity of conducting polymers (more generally, of non-crystalline solids) as a function of frequency consists of a frequency independent low-frequency region and, above a critical frequency value $\omega_c$, a high-frequency one, which is dispersive. Ac conductivity measurements can profile electric charge flow along distances equal or longer proportional to the inverse of the working frequency.

Polaron density exhibits spatio-temporal variations when harmonic electric field of sweeping frequency is applied, because of the inhomogeneity of the disorder. The variation of the frequency influences the charge motion at different spatial scale and results in temporal increased polaron density in spatially restricted territories, such as the conductive grains. The scope of this work is to investigate the above-mentioned hypothesis and the subsequent assertion that the locally increased polaron density in such regions, urge them to undergo a polaron to bipolaron transition.

A complex impedance measurement is performed in the frequency domain at discrete frequency values spanning over a broad frequency range. In this sense, a sweeping frequency of a harmonic electric field, which is applied to a disordered media results in the re-distribution of the polaron population over the inhomogeneous matrix. The variation of the working frequency results in a subsequent variation of the length scale the polarons move. Polarons moving within the volume of the conductive grains yield a dynamic increase of the polaron density in restricted volume space and have increased probability for mutual interaction and formation of bipolarons, which are more stable than polaron. On increasing the ac frequency, polarons move along localized regions; when the characteristic length gets comparable or less than the dimensions of the ordered (conducting) grains, the local increase of polaron concentration is favored and transition is more probable to occur. Such transformation does not alter the net electric charge of the entities contributing to the conductivity, but enhances the electric charge flow due to the resulting bipolaron energy bands. Recalling that the spin of polarons is ½ while bipolarons are spinless, the above-mentioned polaron to bipolaron transformation alters the magnetic susceptibility of the material, as well.

Ac conductivity measurements were performed at room temperature from $10^{-2}$ Hz to 10 MHz by employing a Novo Control Solartron SI 1260 frequency response analyzer. The amplitude of the electric field was 1.5 Volt/mm. We worked on three different materials: (i) teflon, which is actually the test material provided by the manufacturer of our complex impedance analyzer, (ii) sandstone, which is a heterogeneous polycrystalline silicate material with ionic-type electrical conductivity that was extensively studied in previous publications [23 – 26] and (iii) conducting polypyrrole (details about the specimens are given in Ref. [27]). For the first two specimens, successive scans of the real part of the conductivity vs. frequency yielded identical experimental plots. On the contrary, spectacular effects were observed in conducting polypyrrole. The fact that teflon does not exhibit the peculiar one observed in polypyrrole, excludes the possibility of any instrumental artifact. Moreover, the ordinary behavior of sandstone indicates that the dynamic picture of the real part of the conductivity of polypyrrole is related with the nature of the charge carriers. The possibility of electrode polarization is excludes, because: (i) the low frequency conductivity values should decrease in successive experiments due to the space charge limited motion; what is observed is the opposite; i.e., increase (ii) the phenomenon reported is not sensitive to the type of electrodes and the quality of the sample-electrode contact. In Figure 1 successive measurements, recorded during one day, of the real part of the conductivity vs. frequency are depicted. We observe that, at any frequency, the measured conductivity increases after each measurement, with gradually decreasing rate to saturation. Bipolarons are spinless, while polarons have spin ½. If the suggested interpretation for polaron to bipolaron transformation is correct, a systematic change of the magnetic properties of polypyrrole should be observed. The magnetic inductivity L, which is proportional to the magnetic permeability $\mu$, measured at 1.43 MHz corresponding to each measurement is shown in the inset diagram of Figure1. Indeed, we observe a systematic decrease of L in consecutive measurements.

We define a 'bunch' of impedance measurements as a set of conductivity vs. frequency scans collected within one day. Different bunches are also collected at different days. In Figure 2, the first and the final measurement of each bunch at different days are displayed. The general trend is the increase of the conductivity throughout the series of those experiments, while the phenomenon yields saturation values. When reaching to saturation, a back and forth phenomenon appears: the first measurement of a bunch exhibits conductivity values somewhat lower than those detected in the final scan of the previous bunch (Fig. 3).

During the doping process of polypyrrole, polarons have the preference to reduce to bipolarons [28, 29]: a couple of polarons transforms to a bipolaron by reducing a net amount of energy of about 0.45 eV [30]. It is also worth noticing that transverse polarons and bipolarons and their importance for inter-chain conductivity might be related with their stability [31]. We assume that, during a frequency sweep of a measurement, part of the polarons combine to bipolarons. This transformation is intrinsic, in the sense that the specimen is not chemically doped any more. The increased values of the conductivity in successive scans, have two constituents: (i) the transformation of electric charge carriers and (ii) the subsequent energy release in the form of additional phonon energy. Thus, additional bands appear within the energy gap and, moreover, phonons of higher energy are available to contribute to the phonon-assisted hopping. The experimental results that are interpreted through the above mentioned model are:

(i) the fact that the magnetic inductance change in less drastically rate than conductivity does: the former depends solely from the quantity of the transformed electric charge carriers, while in the latter phenomenon additionally contributes the energy released from this transformation,

(ii) the asymptotic increase of the conductivity to saturation, since the polaron population available to coexist in restricted (conductive) regions and are likely to transform to bipolarons, gets successively lower,

(iii) the fact that, when the waiting time between successive measurements is long enough, the conductivity value is slightly lower than the last one measured in a set of successive scans, due to the system relaxation during the waiting time interval by dissipation of the energy released during the transition.

**Figure Captions**

Figure 1: Successive measurements of the real part of the ac conductivity of conducting polypyrrole vs. frequency. Inset diagram: The time sequence is from the bottom curve to the top one. The magnetic inductivity measured at 1.43 MHz, is depicted as a function of the time sequence of the measurements.

Figure 2: The first and the final scan performed within one day for different days (sequent from (a) to (e)). An un-primed letter denotes the first measurement, while a primed on the last one during within the same day. Intermediate measurements were omitted for reasons of simplicity. Part of the diagram is augmented so as to distinguish between curves collapsing to saturation.

Figure 3: The inductivity L measured at a single frequency f=1.43 MHz for different bunches of measurements. A bunch of measurements is a set of impedance scans vs. frequency collected within one day.
Squares, triangles, circles and open triangles correspond to successive 'bunches' of measurements.

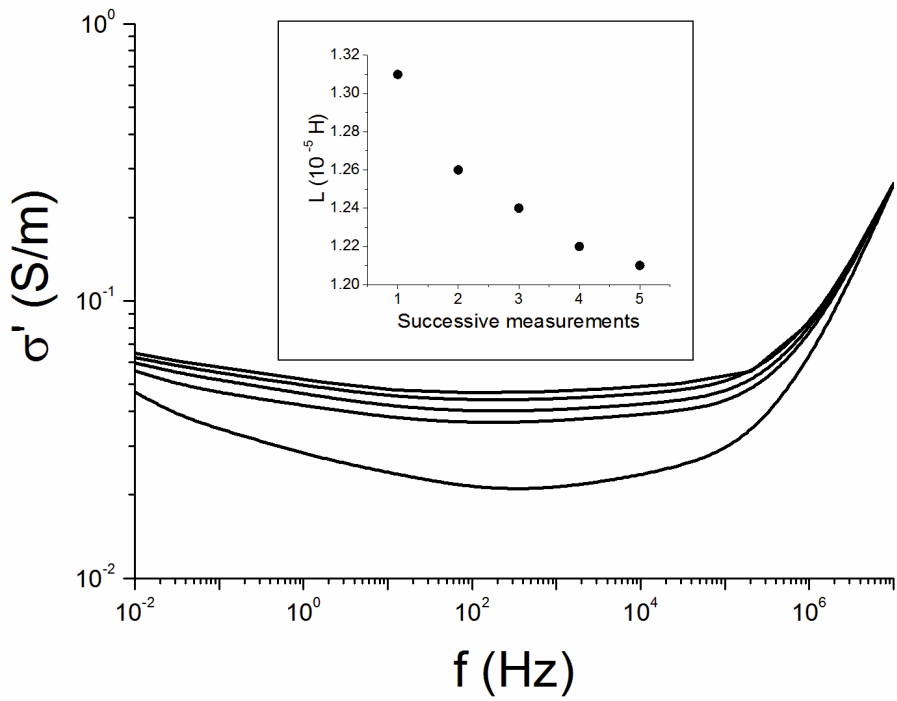

FIGURE 1

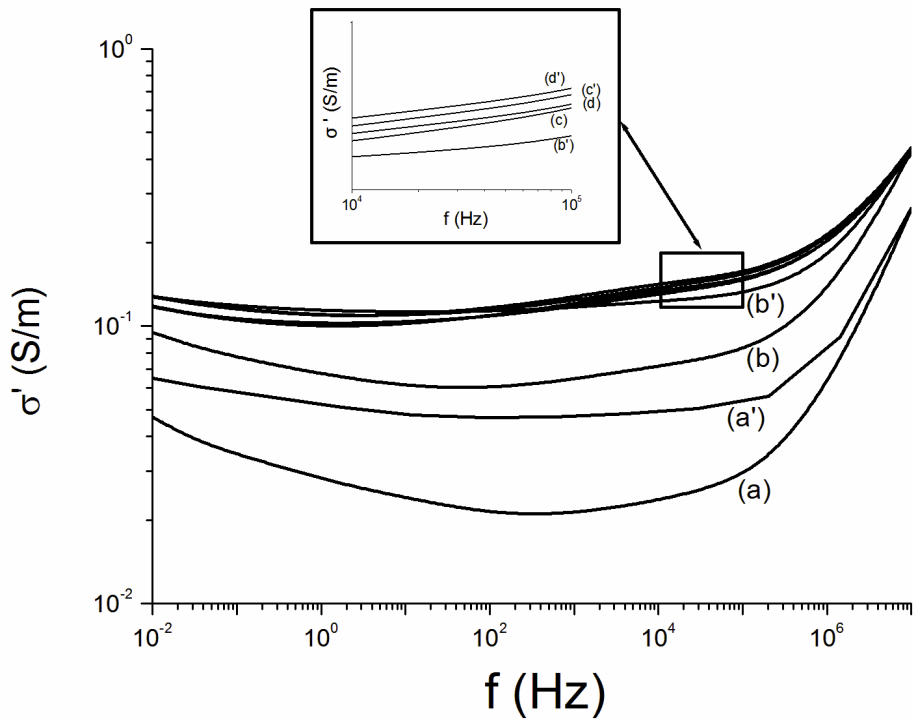

FIGURE 2

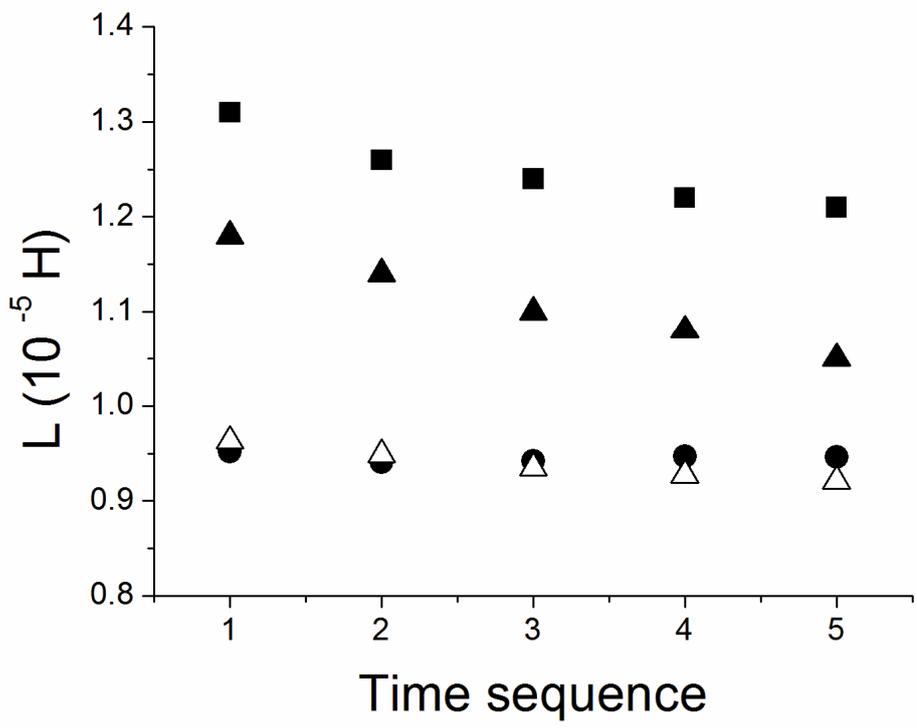

FIGURE 3